\documentclass[twocolumn,10pt]{article}
\usepackage{graphicx}
\usepackage{cite}
\pdfoutput=1
\usepackage{sectsty} 
\allsectionsfont{\large}

\title {Imaging Oxygen Defects and their Motion at a Manganite Surface}
\author {B. Bryant$^1$, Ch. Renner$^2$, Y. Tokunaga$^3$, Y. Tokura$^{3,4,5}$, G. Aeppli$^1$}
\date{\today}
\begin{document}
\maketitle

\footnotetext[1]{London Centre for Nanotechnology and Department of Physics and Astronomy, University College London, London WC1E 6BT, UK}
\footnotetext[2]{Department of Condensed Matter Physics, University of Geneva, 24 Quai Ernest-Ansermet, CH-1211 Geneva 4, Switzerland}
\footnotetext[3]{Multiferroic Project, ERATO, Japan Science and Technology Agency (JST), Wako, 351-0198, Japan}
\footnotetext[4]{Cross-Correlated Materials Research Group (CMRG), RIKEN, Advanced Science Institute, Wako, 351-0198, Japan}
\footnotetext[5]{Department of Applied Physics, University of Tokyo, Bunkyo-ku, Tokyo 113-8656, Japan}

\textbf{Manganites are technologically important materials, used widely as solid oxide fuel cell cathodes: they have also been shown to exhibit electroresistance. Oxygen bulk diffusion and surface exchange processes are critical for catalytic action, and numerous studies of manganites have linked electroresistance to electrochemical oxygen migration. Direct imaging of individual oxygen defects is needed to underpin understanding of these important processes. It is not currently possible to collect the required images in the bulk, but scanning tunnelling microscopy could provide such data for surfaces. Here we show the first atomic resolution images of oxygen defects at a manganite surface. Our experiments also reveal defect dynamics, including oxygen adatom migration, vacancy-adatom recombination and adatom bistability. Beyond providing an experimental basis for testing models describing the microscopics of oxygen migration at transition metal oxide interfaces, our work resolves the long-standing puzzle of why scanning tunnelling microscopy is more challenging for layered manganites than for cuprates.}

\newpage 
Mixed-valence manganites, whose basic units are manganese-centred oxygen octahedra, are used extensively in solid oxide fuel cell cathodes\cite{FuelCellCathodeReview, FuelCellCathodeReview2}, and have attracted great attention due to their colossal magnetoresistance\cite{Tokura:CMR:review}, where electrical resistance is a strong function of external magnetic field. Numerous groups have also examined electroresistance in manganite-based devices, where a low field electrical resistance is switched by the imposition of a higher field\cite{Baikalov:PCMO, sawa:SPIE, Nian:PCMO, Sawa:Review, Shono:PCMO, Harada:PCMO, Odagawa, Tokunaga:APL, Asanuma:PCMO:switching}. Electroresistance has the potential to revolutionise microelectronics through the introduction of new components such as resistive random access memory cells\cite{Sawa:Review} and memristors\cite{Chua:Memristor,Strukov:memristor}. Electroresistance has been attributed, by several groups, to the electrochemical migration of oxygen ions\cite{Shono:PCMO, Sawa:Review, Baikalov:PCMO, Nian:PCMO, Asanuma:PCMO:switching}. Diffusion of oxygen through the bulk material and at the surface  are also crucial processes for catalytic action in fuel cell cathodes\cite{FuelCellCathodeReview2}. Direct imaging of oxygen defects, and of surface oxygen diffusion would be a distinct advantage for the understanding of these processes: time resolved scanning tunnelling microscopy could in principle provide such images.

A key parameter in manganite materials is the formal valence state of the manganese ions, which is tuned by varying the mix of divalent and trivalent cations. Electroresistance is strongest close to 50\% divalent ion doping\cite{Tokunaga:APL, Asanuma:PCMO:switching}, where the formal Mn valence is a half integer. Likewise, catalytic activity in manganites is enhanced by mixed valency\cite{FuelCellCathodeReview}. Oxygen defects such as excess oxygen ions and vacancies, implicated in both catalysis and electroresistance, are also most probable at half doping. Therefore, to maximise our ability to image oxygen defects in a manganite using scanning tunnelling microscopy (STM), we require a half-doped manganite, whose surface can be prepared reliably. Unlike the three-dimensional pseudocubic manganites, layered manganites include planes along which the material can be cleaved, producing an atomically flat surface. Accordingly, we have chosen to study PrSr$_2$Mn$_2$O$_7$, a half doped bilayered manganite built from double sheets (figure \ref{Figure1}a) of pseudocubic (Pr,Sr)MnO$_3$ alternating with sheets of rock-salt-like (Pr,Sr)O. 

STM studies of manganites are uncommon, particularly when compared to those of the closely related high temperature superconducting copper oxides. Previous experiments have demonstrated clear atomic resolution\cite{Renner:BCMO, Ma:LPCMO, Fuchigami, Roessler} for pseudocubic manganites, but atomically resolved STM studies of layered manganites are rare\cite{renner:LSMO}, despite the fact that the layered structure provides cleaving planes. 

In this paper, we report atomic resolution STM imaging of the bilayered manganite PrSr$_2$Mn$_2$O$_7$. As well as atomic resolution on the atomically flat surface, oxygen defects, in the form of both adatoms and vacancies, are observed at the PrSr$_2$Mn$_2$O$_7$ surface. Time resolved imaging has been used to demonstrate oxygen adatom migration and vacancy-adatom recombination: voltage dependent oxygen motion and adatom bistability are also observed. Our results bring the benefits of atomic resolution to the study of manganite fuel cell cathode materials and electroresistive devices.

\section*{Results}

\begin{figure}
\begin{centering}
\includegraphics[width=0.5\textwidth]{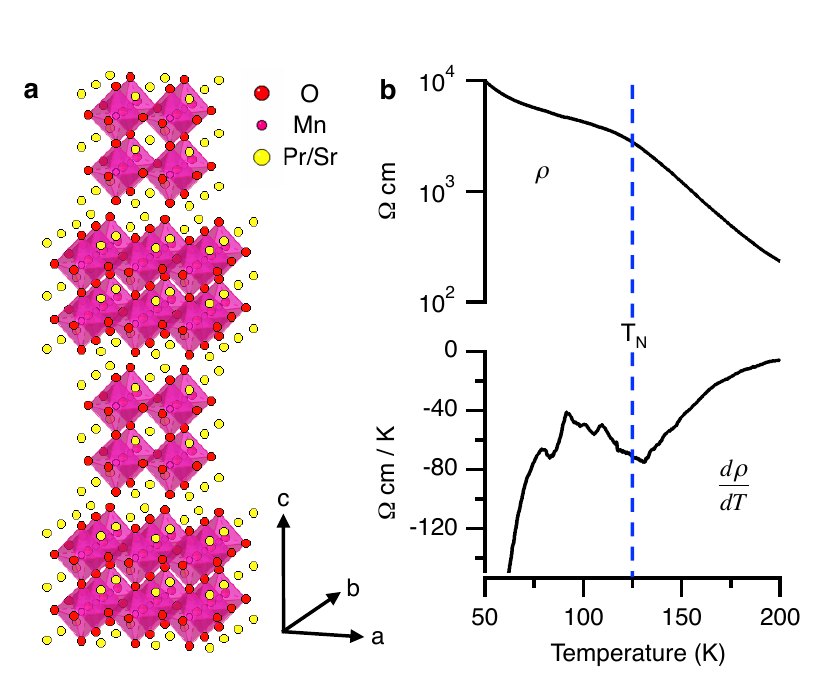}
\caption{\label{Figure1} \textbf{Structure and properties of PrSr$_2$Mn$_2$O$_7$.} (a) Crystal structure of PrSr$_2$Mn$_2$O$_7$, showing bilayered structure. (b) c-axis resistivity $\rho$ and differential resistivity $d\rho/dT$ of PrSr$_2$Mn$_2$O$_7$ versus temperature. The antiferromagnetic ordering temperature T$_N$ $=$ 125 K is indicated.}
\end{centering}
\end{figure}

\begin{figure*}[p]
\begin{centering}
\includegraphics[width=0.9\textwidth]{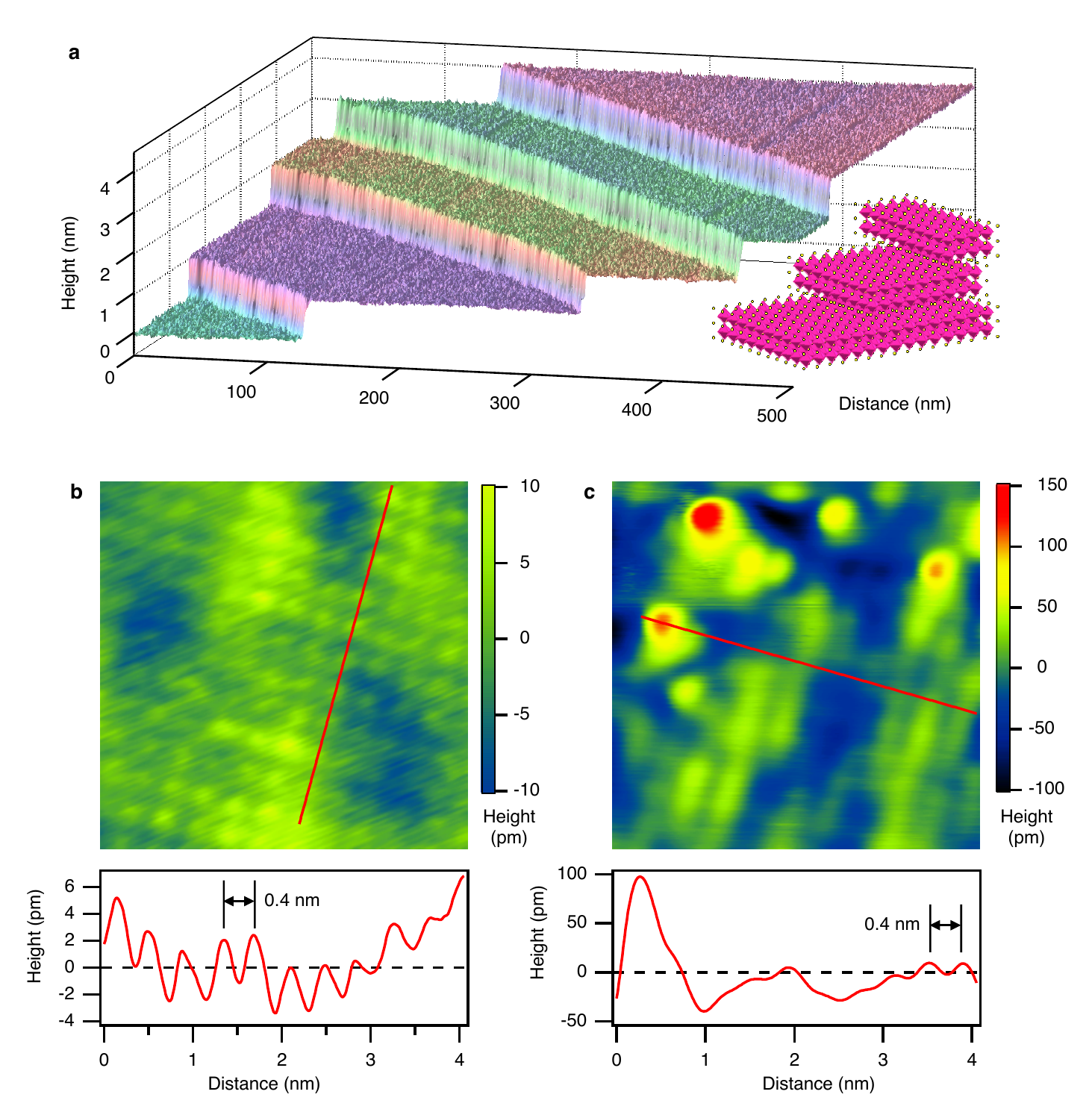}
\caption{\label{Figure2} \textbf{STM microscopy of PrSr$_2$Mn$_2$O$_7$}. (a) 500 x 500 nm$^2$ micrograph of PrSr$_2$Mn$_2$O$_7$\ collected at 78 K, 100 pA tunnel current and +0.8 V sample bias. The observed step height of 1.01 $\pm$ 0.02 nm is consistent with c/2 = 0.996 nm\cite{Tokura:PSMO}. (b) 4 x 4 nm$^2$ micrograph (125 K, 50 pA, -0.5 V) revealing a weak atomic lattice contrast, superposed on a 3-4 nm length scale inhomogeneity. A section through the micrograph is shown: the atomic lattice has $\approx$ 4 pm peak-to-peak corrugation. (c) 4 x 4 nm$^2$ micrograph (78 K, 100 pA, -0.8 V) showing adatoms, vacancies and atomic rows: a section through an adatom is shown.}
\end{centering}
\end{figure*}

\begin{figure*}[t]
\begin{centering}
\includegraphics[width=0.9\textwidth]{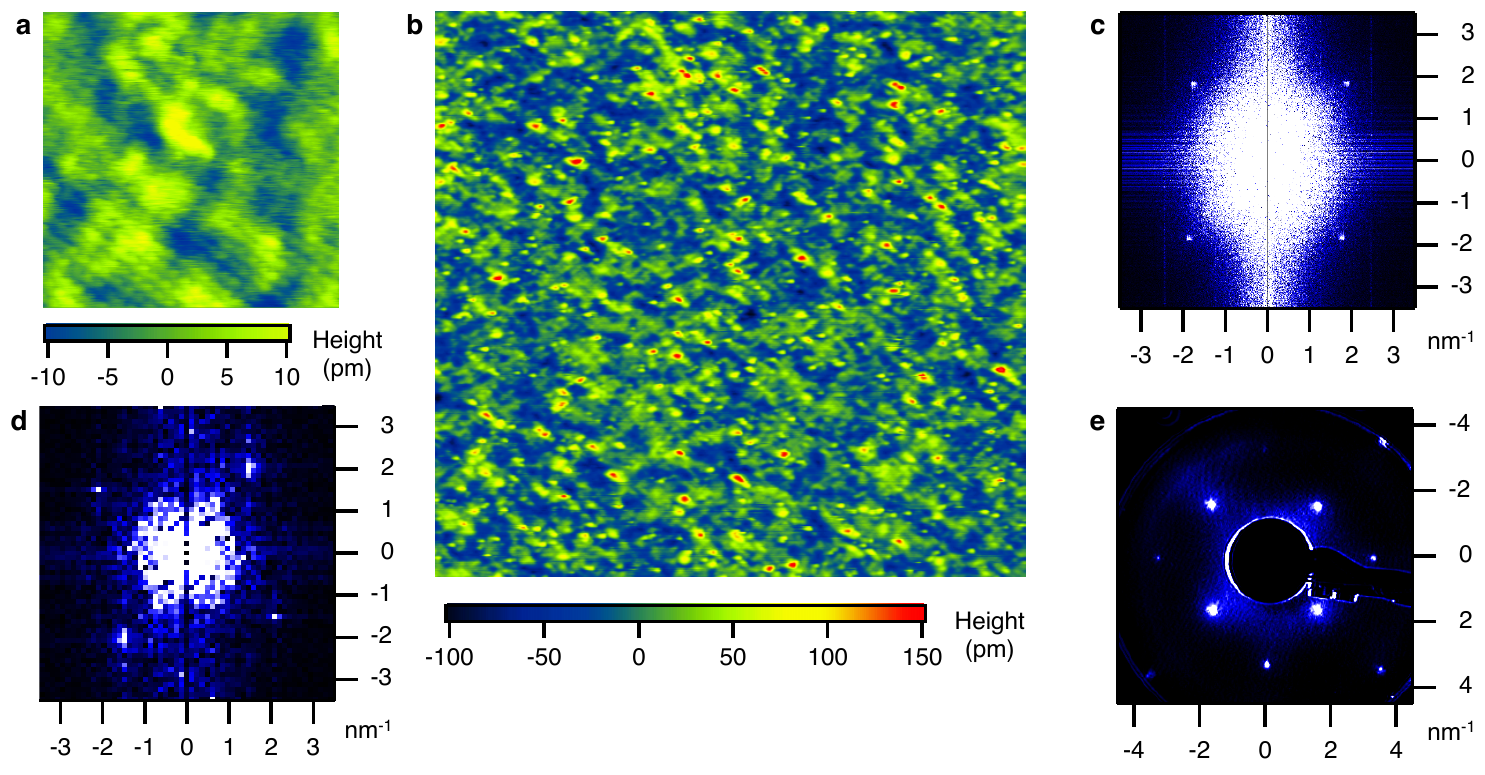}
\caption{\label{Figure3} \textbf{STM and LEED data.} (a) 8 x 8 nm$^2$ micrograph (78 K, 50 pA, +0.5 V) showing atomic lattice and surface inhomogeneities. (b) 50 x 50 nm$^2$ micrograph (78 K, 100 pA, -0.8 V) showing adatoms and vacancies. (c) Fourier transform of (b), showing peaks due to square lattice of spacing a = 0.382 $\pm$ 0.003 nm, corrugation 0.8 $\pm$ 0.1 pm. (d) Fourier transform of (a) showing peaks due to square lattice of spacing a = 0.40 $\pm$ 0.01 nm, corrugation 3.8 $\pm$ 0.2 pm. (e) LEED image from the same surface as (c), with the sample in the same orientation, showing peaks due to a square lattice, spacing a =  0.43 $\pm$ 0.03 nm.}
\end{centering}
\end{figure*}

\textbf{Atomic resolution STM imaging of PrSr$_2$Mn$_2$O$_7$.} PrSr$_2$Mn$_2$O$_7$ is semiconducting, and is antiferromagnetic below T$_N$ = 125 K\cite{Tokura:PSMO}. Figure \ref{Figure1}b shows the c-axis resistivity and differential resistivity of a PrSr$_2$Mn$_2$O$_7$ single crystal, as used for STM experiments: an inflection is seen in the resistivity at around T$_N$. Figure \ref{Figure2}a is an STM topograph of cleaved PrSr$_2$Mn$_2$O$_7$, together with the crystal structure. Terraces several hundred nm across are visible, separated by steps of height $1.01 \pm 0.02$ nm, corresponding to c/2 = 0.996 nm as measured by X-ray diffraction\cite{Tokura:PSMO}. Atomic resolution can be achieved right across these terraces for this poor electrical conductor, in contrast with previous studies on La$_{1.4}$Sr$_{1.6}$Mn$_{2}$O$_{7}$\cite{renner:LSMO}, which is a much better metal and where screening of atoms is more effective. Figure \ref{Figure2}b shows a high resolution STM topograph, displaying a square atomic lattice with a = 0.40 $\pm$ 0.01 nm. This is in reasonable agreement with the value from X-ray diffraction of a = 0.385 nm\cite{Tokura:PSMO}. Figures \ref{Figure2}b and  \ref{Figure3}a reveal, in addition to the atomic lattice, surface inhomogeneities with $\approx$ 20 pm corrugation. The latter depend on bias polarity (see supplementary figure S1) and therefore have an electronic component, most likely due to charge inhomogeneities (mixed Mn valence and/or cation distribution). These features are similar to the surface inhomogeneities observed on the La$_{0.5}$Sr$_{1.5}$MnO$_4$ surface by STM\cite{Plummer2,Takagi} and surface x-ray scattering\cite{Wakabayashi}.

For the isostructural compound La$_{2-2x}$Sr$_{1+2x}$Mn$_2$O$_7$, the surface exposed on cleaving has been identified as a (La,Sr)O plane, based on a combination of crystal structure arguments, STM imaging and x-ray photoelectron spectroscopy\cite{Loviat}. In our STM images of PrSr$_2$Mn$_2$O$_7$\, all terrace steps heights are integer multiples of c/2: no variation in step height is observed. Two possible mirror planes may yield a step height of c/2, bisecting either a single perovskite bilayer, or a rock salt layer. The bilayer is equivalent in structure to a perovskite manganite: these are known not to cleave naturally\cite{Renner:BCMO}. The cleaving plane through the rock-salt layer is energetically favoured\cite{Loviat}, implying that the observed surface is a (Pr,Sr)O layer. 

\textbf{Imaging of adatoms and vacancies at the PrSr$_2$Mn$_2$O$_7$\ surface.} Certain areas of the PrSr$_2$Mn$_2$O$_7$\ surface show a much increased roughness of around 250 pm peak to peak. Figure \ref{Figure2}c shows a high-resolution STM topograph of such an area while figure \ref{Figure3}b shows a larger area image. The increased roughness is derived from a distribution of raised ``adatoms'' and depressed  ``vacancies''. The adatom coverage is variable, with a maximum around one per 65 surface unit cells. We may rule out the possibility that this rougher surface represents a different cleaving plane to the atomically flat surface, since terrace step heights are still restricted to multiples of c/2. Terrace steps of distinct heights, as observed on the single-layered manganite La$_{0.5}$Sr$_{1.5}$MnO$_4$\cite{Takagi}, indicating different surface terminations\cite{Wakabayashi}, were not seen for  PrSr$_2$Mn$_2$O$_7$. 

\begin{figure*}[t]
\begin{centering}
\includegraphics[width=0.9\textwidth]{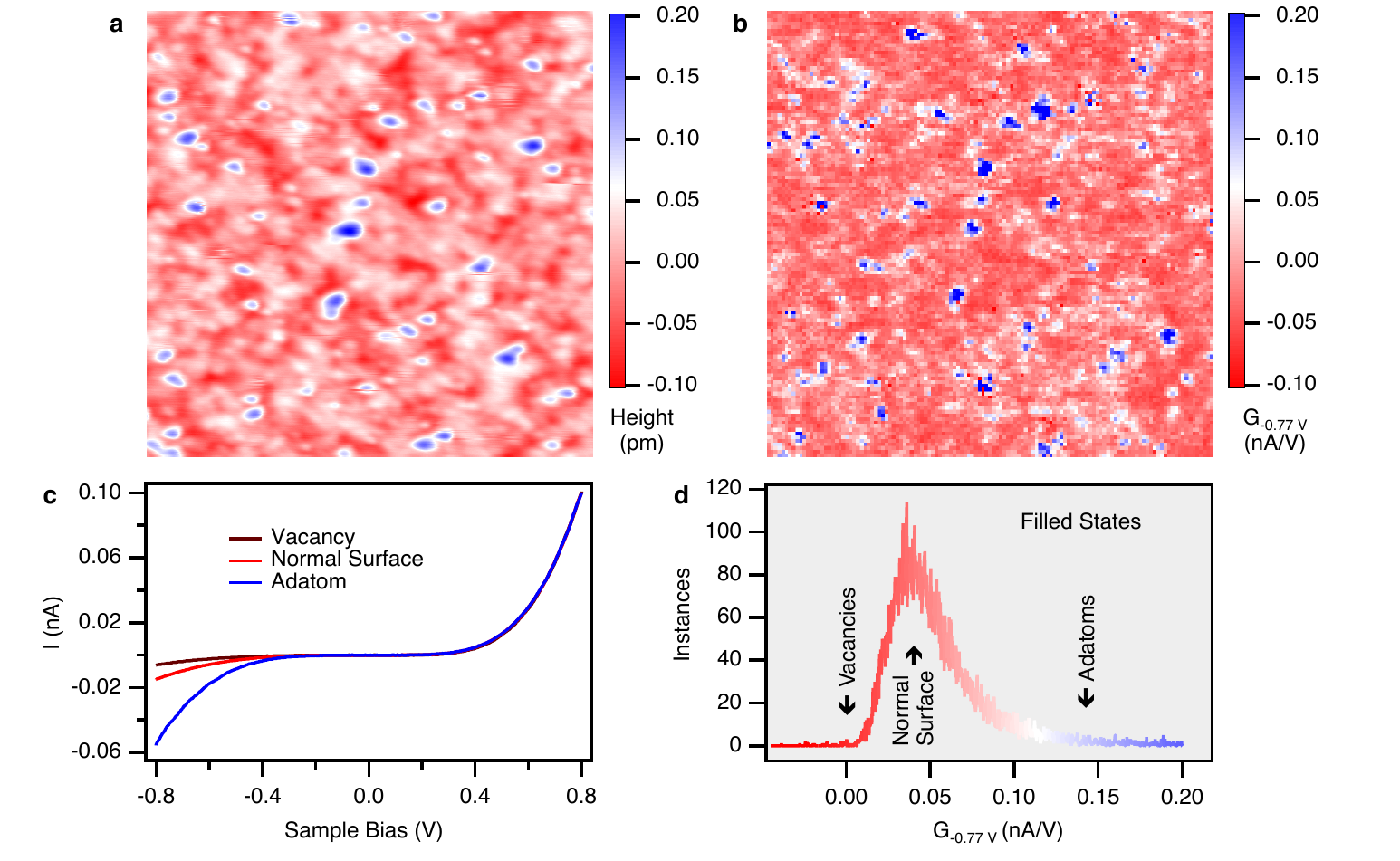}
\caption{\label{Figure4} \textbf{STS data from PrSr$_2$Mn$_2$O$_7$.} (a) 20 x 20 nm$^2$ STM micrograph collected at 140 K (-0.8 V, 100 pA). (b) Conductivity map derived from STS map acquired simultaneously for the same area, displayed at -0.77 V. (c) Example spectra of adatoms, vacancies and the normal PrSr$_2$Mn$_2$O$_7$ surface. (d) Histogram of (b), with the same colour scale, showing the conductivity associated with adatoms, vacancies and the normal surface.}
\end{centering}
\end{figure*}

On areas of the PrSr$_2$Mn$_2$O$_7$\ surface showing adatoms and vacancies in STM images, it is more difficult to achieve good atomic resolution than on the atomically flat surface. Nonetheless, atomic rows are visible in figure \ref{Figure2}c: Fourier transforms of larger scan areas (figure \ref{Figure3}c) show peaks corresponding to the same square lattice as the atomically flat surface (figure \ref{Figure3}d), with measured lattice constant a = 0.382 $\pm$ 0.003 nm. The same lattice is thus observed on both atomically flat and adatom-vacancy areas. No consistent ordering of adatoms was observed using either Fourier or self-correlation image analysis: the adatoms appear randomly distributed on the surface, with no reference to the atomic lattice. 

The surface atomic lattice can also be observed via low energy electron diffraction (LEED). Figure \ref{Figure3}e shows the LEED image of the same cleaved surface as in figure \ref{Figure3}b; the sample  orientation is preserved upon translation from the STM chamber to the LEED/sample preparation chamber while maintaining ultra-high vacuum. The LEED peaks correspond to a square lattice with a =  0.43 $\pm$ 0.03 nm, indistinguishable from the STM value. LEED measurements have been carried out at a range of temperatures from 20 K to 300 K, but no ``satellite peaks'', indicative of surface charge ordering, appeared at any temperature.

\textbf{Scanning tunnelling spectroscopy of adatoms and vacancies.} We mapped spatial variations of the electronic density of states using scanning tunnelling spectroscopy (STS). Figure \ref{Figure4}a shows an STM topograph of an area of the PrSr$_2$Mn$_2$O$_7$ surface showing adatoms and vacancies. Figure \ref{Figure4}b shows a conductance map, which has been derived from an STS map acquired simultaneously with the topograph. The conductance is displayed in filled states at a sample bias voltage of -0.77 V. There is a clear correlation between adatoms visible in the topographic image, and peaks in the filled state conductance. Figure \ref{Figure4}c shows example spectra of adatoms, vacancies and the normal PrSr$_2$Mn$_2$O$_7$ surface. Adatoms are associated with an excess of filled state tunnel current (representing integrated density of states) over the normal surface, vacancies with a slight deficiency. 

\begin{figure}
\begin{centering}
\includegraphics[width=0.43\textwidth]{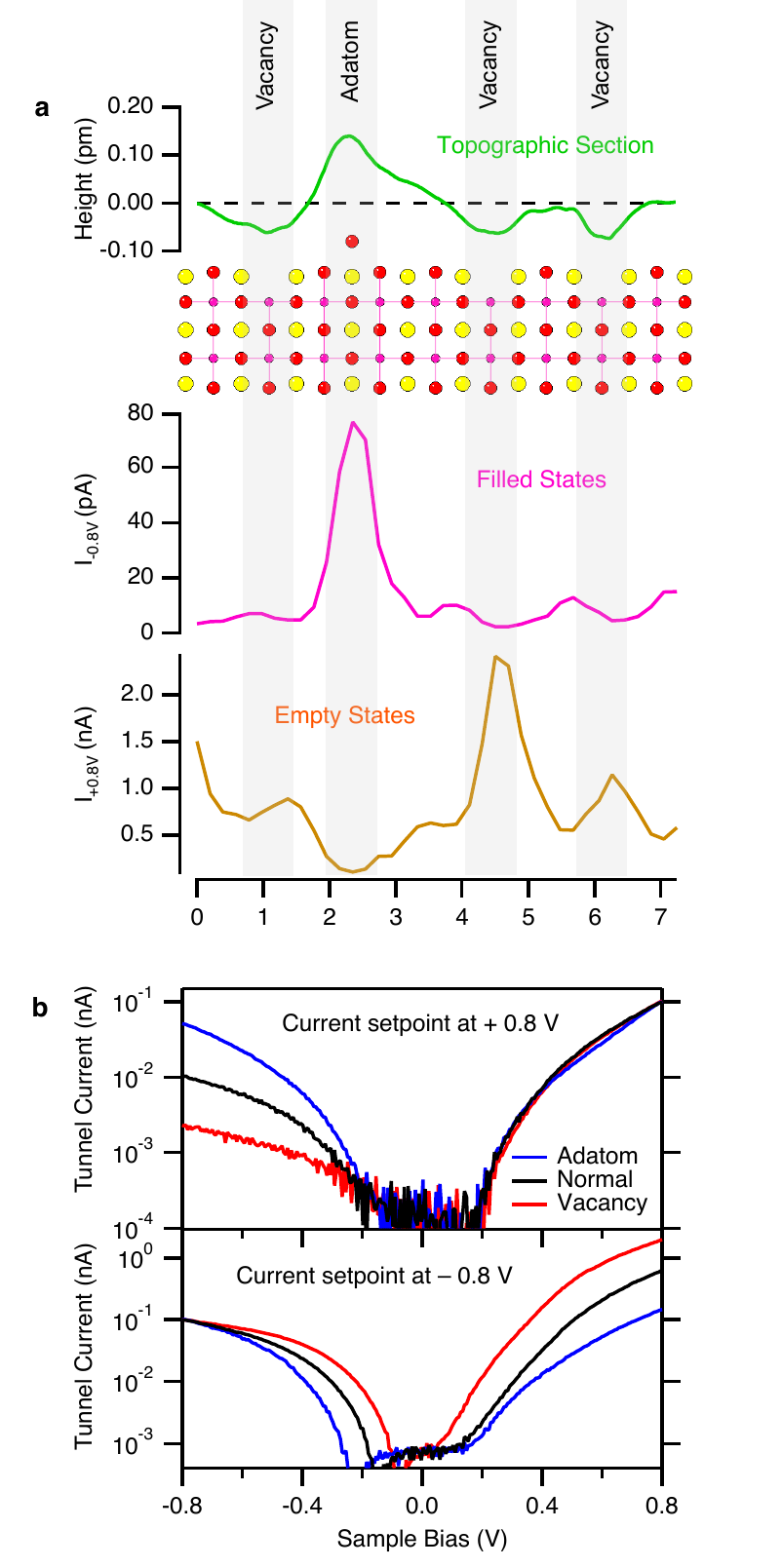}
\caption{\label{Figure5} \textbf{Nature of adatoms and vacancies.} (a) Topographic section of an increased roughness area of the PrSr$_2$Mn$_2$O$_7$\ surface (-0.8 V, 100 pA), and the proposed model with oxygen adatoms and vacancies. Also shown are sections though two I(V) maps at the same position, showing the magnitude of the tunnel current in both filled (-0.8V) and empty (+0.8V) states, where tip lift heights z were fixed for 100 pA tunnel current at +0.8 V and -0.8 V respectively. (b) Sample spectra from these I(V) maps. All data in (a) and (b)  were collected at 78 K.}
\end{centering}
\end{figure}

Figure \ref{Figure5}a shows a topographic section, which includes three vacancies and an adatom, and sections through two I(V) maps, at -0.8V (filled states) and +0.8V (empty states). The magnitude of the tunnel current, which represents the integrated density of states, is plotted along the same x-axis used for the topographic section. The filled state current shows a large peak at the position of the adatom, while the empty state current shows peaks at the positions of the vacancies. Figure \ref{Figure5}b shows example I(V) spectra from adatoms, vacancies and the normal surface, from both I(V) maps. It is clear that the adatom spectra have a large excess of tunnel current at negative bias (filled states) compared to the normal surface, and that the tunnel current from vacancies is suppressed, while for positive bias (empty states) the reverse is true. Based on these data and the expectation that the surface is a (Pr,Sr)O layer, we interpret the bright adatoms seen in topographic images as oxygen, and the dark depressions as oxygen vacancies. This explains the large excess in filled states seen at adatom sites, as the oxygen will be negatively charged. Analagously, oxygen vacancies will carry a positive charge.  A similar effect was seen in I(V) spectra of La$_{5/8}$Ca$_{3/8}$MnO$_3$, where an incomplete oxygen overlayer was formed on the MnO$_2$ terminated surface, by annealing in an oxygen atmosphere\cite{Fuchigami}. Since adatoms and vacancies are observed when the sample has been freshly cleaved, and the density of adatoms is not seen to increase, even after two weeks in the UHV chamber, we suggest that the oxygen adatoms and vacancies appear not as a result of deposition or contamination, but that they arise when the sample is cleaved. Some oxygen ions may be removed with the cleaved-off part of the sample, forming surface vacancies. Similarly, oxygen ions from the cleaved-off part may be left on the surface as adatoms. Figure \ref{Figure5}a represents the proposed model together with the topographic section; the [100] plane is shown, with Pr/Sr sites in yellow and O in red. The model is plausible since adatoms and vacancies occur in roughly equal numbers, as expected by symmetry. Contrast reversal is not observed between empty and filled state images (see supplementary figure S2), indicating that the ``adatoms'' and ``vacancies'' are not purely electronic contrast features but have physical height and depth. 

\begin{figure*}[t]
\begin{centering}
\includegraphics[width=1.0\textwidth]{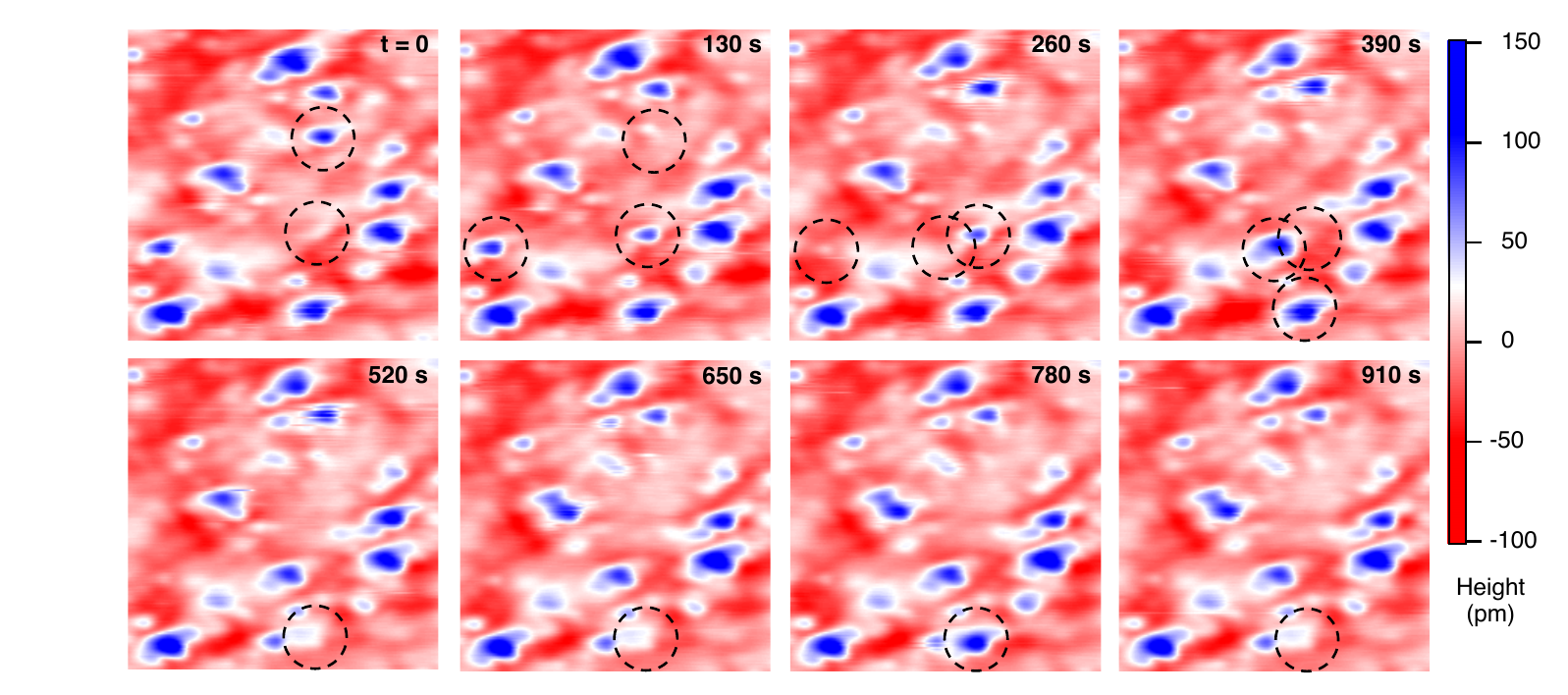}
\caption{\label{Figure6} \textbf{Adatom motion.} Time series of filled state STM micrographs,  10 x 10 nm$^2$, collected at 170 K (-0.8 V, 100 pA). The time per frame is 130 s: the total time elapsed is 910 s. Some adatoms (circled in adjacent frames) move from one frame to the next. See also supplementary movie 1.}
\end{centering}
\end{figure*}

\begin{figure*}[t]
\begin{centering}
\includegraphics[width=1.0\textwidth]{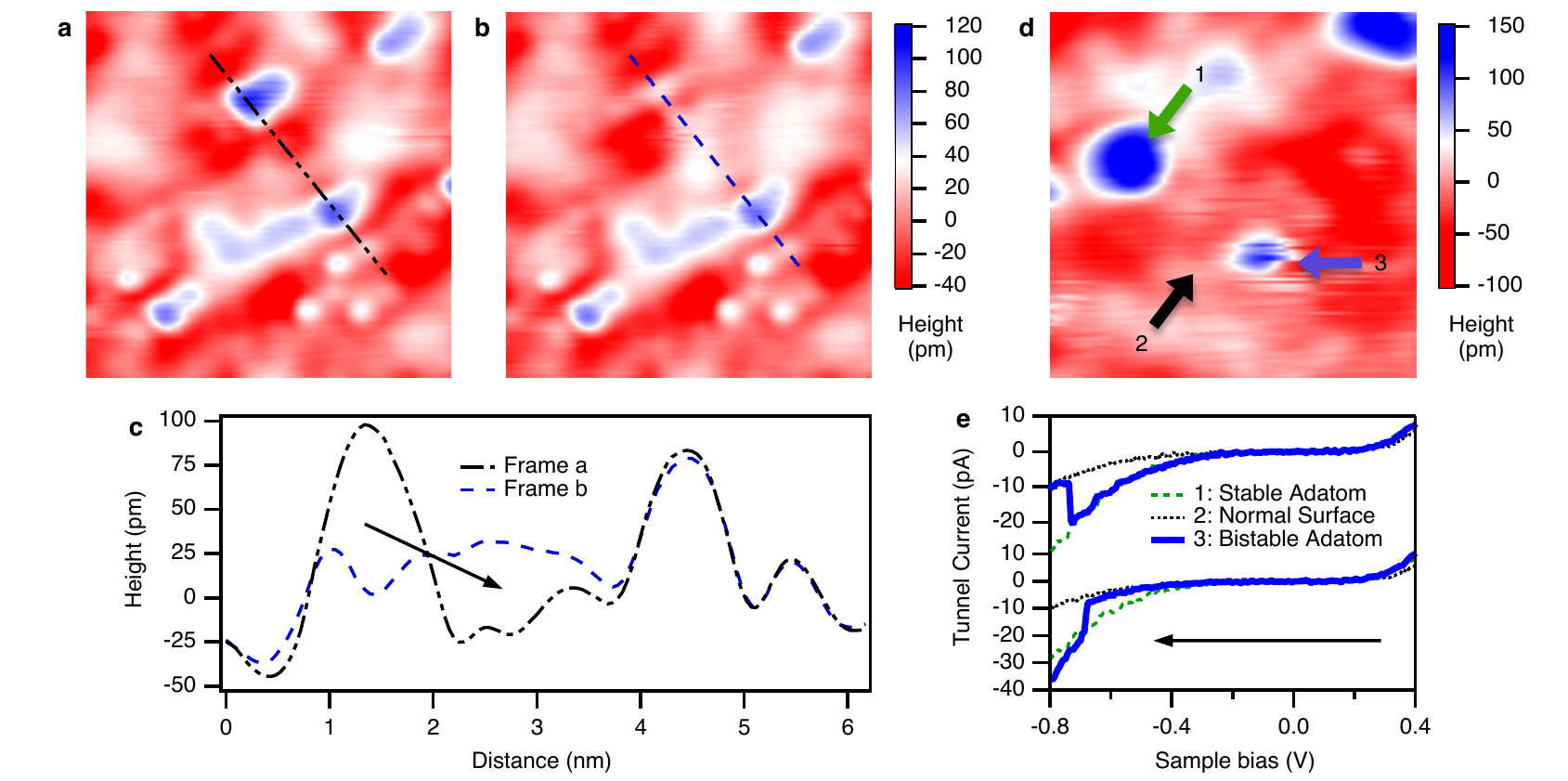}
\caption{\label{Figure7} \textbf{Recombination of adatoms and vacancies.} (a) and (b): successive frames in a time series showing recombination of an adatom/vacancy pair. Frames are 8 x 8 nm$ ^2$ and are separated by 1 minute: images collected at 170 K (100 pA, +0.8V). (c) Line profiles through (a) and (b) showing that both the adatom and the adjacent vacancy vanish. (d) 6 x 6 nm$^2$ STM topograph showing a stable adatom (1) and a bistable adatom (3). The adatom appears streaky because it moves laterally between one scan line and the next: the STM tip takes around 10 s to scan over the adatom. Image collected at 125 K (100 pA, -0.8). (e) I(V) spectra at these positions, with a spectrum from the normal surface (position 2). Two spectra are shown from the bistable adatom: these show the characteristic discontinuity caused by motion of the adatom. The spectra can be seen to switch between that typical of an adatom and that of the normal surface. The voltage is swept from positive to negative.}
\end{centering}
\end{figure*}

\textbf{Oxygen adatom motion and adatom-vacancy recombination.} The oxygen adatom positions are not all fixed. Figure \ref{Figure6} (and supplementary movie 1) shows a time series of STM images where several adatoms on the PrSr$_2$Mn$_2$O$_7$\ surface appear or disappear between frames, although most adatoms remain stationary. We interpret this as hopping of adatoms from site to site, rather than single-site switching since the appearance and disappearance of adatoms are correlated. Indeed, we sometimes observe cascades where adatoms appear to sequentially hop from one site to another. In figure \ref{Figure6} it may be observed that the adatom site vacated at 520 s is re-occupied at 780 s, and then vacated again at 910 s. This effect has been observed multiple times and implies that adatoms have preferred positions  on the PrSr$_2$Mn$_2$O$_7$\ surface. Since adatoms are not observed to form ordered arrays on the surface, these preferred positions are not determined by the atomic lattice. It is more likely that the surface inhomogeneities seen in figures \ref{Figure2}b and \ref{Figure3}a, which in turn are likely due to charge inhomogeneities (mixed Mn valence and/or distribution of Pr$^{3+}$ and Sr$^{2+}$ ions), describe a potential landscape upon which oxygen adatoms move. The observed mobility provides further evidence that these features are surface adatoms. Adatom mobility is strongly dependent on sample bias polarity, with negligible mobility at positive sample bias, suggesting an electrostatic component to the hopping mechanism. Mobile adatoms have been observed at a range of temperatures, with the adatom mobility increasing with sample temperature (see supplementary figure S3, and supplementary discussion). The total number of adatoms and vacancies per image decreases with increasing sample temperature (supplementary figure S3). We attribute this to recombination of oxygen adatoms and vacancies, a  process which we have  observed directly. For example, figure \ref{Figure7} shows two successive frames of a time series. In (a) an adatom and a vacancy are visible in close proximity, while in (b) both the adatom and the vacancy have vanished, suggesting that the adatom has filled the vacancy. This can most clearly be seen in the comparative section (figure \ref{Figure7}c). The observation of oxygen adatom motion provides a way to directly measure the surface oxygen exchange coefficient, a crucial parameter for the oxygen reduction action of solid oxide fuel cell cathodes\cite{FuelCellCathodeReview, FuelCellCathodeReview2}. Similarly, our observation of adatom-vacancy recombination demonstrates how adsorbed oxygen atoms may be incorporated into the cathode material.

\textbf{Bistable oxygen adatoms.} In addition to adatom hopping observed in time series of images, bistable adatoms appear in single (rastered) STM images and I(V) spectra. Such adatoms can switch between two adjacent positions several times during a single topographic scan. Figure \ref{Figure7}d is an STM topograph including a bistable adatom, which moves laterally on the surface during the scan, resulting in a ``streaky'' image of the adatom. Figure \ref{Figure7}e shows two I(V) spectra collected at the same bistable adatom. These spectra are discontinuous because the adatom switches position during the I(V) measurement: the measured spectra switch between that typical of an oxygen adatom and that of the underlying surface in less than the time resolution of 1 ms. From figure \ref{Figure7}d the lateral movement of the oxygen adatom may be measured as less than 0.2 nm. Bistable I(V) spectra have been identified on many adatoms (see supplementary figure S4), and discontinuities are observed for both sweep directions at a range of positive and negative voltages. Single-atom switching for oxygen on a manganite surface may be compared to switching of xenon atoms on nickel surfaces\cite{Eigler:switch}, and hydrogen on Si[100]\cite{Quaade:atomswitch}. The lateral motion seen for the H adatoms is more similar to our discovery; the xenon moved not across the surface but between the STM tip and the metal surface.

\section*{Discussion}

We have pioneered the visualisation of oxygen defects and their motion for a perovskite oxide of manganese. Our discovery, including the choice of sample and low temperature cleaving, enables future microscopic investigations of manganite fuel cell cathode materials. In such investigations, molecular oxygen would be bled into the STM vacuum chamber and the oxygen incorporation process at the manganite surface imaged, allowing the role of vacancies in oxygen reduction to be observed directly. Our work also resolves the long-standing puzzle of why the stable, atomically flat surfaces needed for scanning tunnelling microscopy are much less common for layered manganites than for cuprates, by showing clearly the surface roughness associated with oxygen defects at a manganite surface. Oxygen adatoms are mobile at temperatures well below room temperature, creating a highly challenging surface for STM investigations.

Oxygen motion has been held responsible for resistance switching in numerous studies of manganites\cite{Shono:PCMO, Sawa:Review, Baikalov:PCMO, Nian:PCMO, Asanuma:PCMO:switching}. Most studies indicate that electroresistance in manganite-based devices occurs at the interface between the manganite and the metal contact rather than in the bulk material\cite{Baikalov:PCMO, sawa:SPIE, Nian:PCMO, Sawa:Review, Shono:PCMO, Harada:PCMO, Tokunaga:APL}. The oxidation and reduction of either the manganite itself\cite{Nian:PCMO}, or a metal oxide layer\cite{Shono:PCMO, Asanuma:PCMO:switching} at the manganite-metal interface, as a result of electrochemical migration of oxygen ions, has been proposed as the underlying mechanism. Our work brings the benefit of atomic resolution to the study of electroresistance in manganites. Voltage dependent oxygen motion and adatom bistability are observed at the interface between the manganite surface and the metal STM tip. We have demonstrated the unprecedented ability, not provided by meso- and macroscopic devices, to view the voltage-induced displacement of single oxygen atoms at a manganite surface.   

\section*{Methods}

\textbf{Sample growth and preparation.} We used standard optical float zone methods to grow rod-like crystals of PrSr$_2$Mn$_2$O$_7$, with typical diameter 5 mm and length 100 mm. PrSr$_2$Mn$_2$O$_7$\ crystals were cleaved in situ in ultrahigh vacuum at 20 K and then immediately loaded into the STM, at 78 K. Cleaving at low temperature allowed reliable preparation of atomically flat and clean surfaces. Six cleaves were performed successfully, on crystals from two batches. 

\textbf{STM experiments.} STM measurements were carried out in ultrahigh vacuum ($<$5x$10^{-11}$ mbar) using electrochemically etched tungsten tips derived from 250 $\mu$m diameter wire. The experiments were conducted in the temperature range 78 K to 195 K. Images and I(V) spectra were typically collected at a tunnel current setpoint of 100 pA and a bias voltage of $\pm$ 0.8 V. The STM piezo scanner was calibrated in three dimensions using topographic images of the Si (100) surface as a reference. Atomically flat PrSr$_2$Mn$_2$O$_7$\ surfaces have been observed at 78 K and 125 K, while surfaces with adatoms and vacancies have been observed at a range of temperatures from 78 K to 195 K. The two types of surface have been observed on the same sample in different locations. Both types of surface have been observed immediately after cleaving the PrSr$_2$Mn$_2$O$_7$\ crystal. 

\textbf{Dual mode STS mapping.} For the data shown in figure \ref{Figure5} dual mode STS mapping has been used to allow two STS measurements with different parameters to be collected simultaneously. The filled state section is derived from an STS map in which the sample bias was swept from positive to negative, and the tunnel current setpoint is set at positive bias: the empty state section is derived from an STS map in which the bias is swept from negative to positive, and the current is set at negative bias. In both cases the current setpoint was 100 pA.

\section*{Acknowledgements}

We thank K. Iwaya for guidance with STM methods, A.M Stoneham and A. Fisher for helpful discussions, and C. F. Hirjibehedin, S. Wirth and H. M. R\o nnow for assistance in preparing the paper. The work was supported by the Engineering and Physical Sciences Research Council of the United Kingdom and the European Union's Framework Six programme.

\providecommand{\href}[2]{#2}\begingroup\raggedright\endgroup


\begin{thebibliography}{10}

\bibitem{FuelCellCathodeReview}
S.~P. Jiang, ``{Development of lanthanum strontium manganite perovskite cathode
  materials of solid oxide fuel cells: a review},''
  \href{http://dx.doi.org/10.1007/s10853-008-2966-6}{{\em {Journal of Materials
  Science}} {\bfseries {43}} no.~{21}, ({Dec}, {2008}) {6799--6833}}.

\bibitem{FuelCellCathodeReview2}
C.~Sun, R.~Hui, and J.~Roller, ``{Cathode materials for solid oxide fuel cells:
  a review},'' \href{http://dx.doi.org/10.1007/s10008-009-0932-0}{{\em Journal
  of Solid State Electrochemistry} {\bfseries {14}} no.~{7}, (July, {2010})
  {1125--1144}}.

\bibitem{Tokura:CMR:review}
Y.~Tokura, ``Critical features of colossal magnetoresistive manganites,''
  \href{http://dx.doi.org/10.1088/0034-4885/69/3/R06}{{\em Reports on Progress
  in Physics} {\bfseries 69} no.~3, (2006) 797--851}.

\bibitem{Baikalov:PCMO}
A.~Baikalov, Y.~Wang, B.~Shen, B.~Lorenz, S.~Tsui, Y.~Sun, Y.~Xue, and C.~Chu,
  ``{Field-driven hysteretic and reversible resistive switch at the
  Ag-Pr$_{0.7}$Ca$_{0.3}$MnO$_3$ interface},''
  \href{http://dx.doi.org/10.1063/1.1590741}{{\em Applied Physics Letters}
  {\bfseries {83}} no.~{5}, ({Aug 4}, {2003}) {957--959}}.

\bibitem{sawa:SPIE}
A.~{Sawa}, T.~{Fujii}, M.~{Kawasaki}, and Y.~{Tokura},
  \href{http://dx.doi.org/10.1117/12.616682}{``{Interface transport properties
  and resistance switching in perovskite-oxide heterojunctions},''} in {\em
  Strongly Correlated Electron Materials: Physics and Nanoengineering},
  I.~{Bozovic} and D.~{Pavuna}, eds., vol.~5932 of {\em Proceedings of the
  SPIE}, pp.~342--349.
\newblock Jan., 2005.

\bibitem{Nian:PCMO}
Y.~B. Nian, J.~Strozier, N.~J. Wu, X.~Chen, and A.~Ignatiev, ``{Evidence for an
  oxygen diffusion model for the electric pulse induced resistance change
  effect in transition-metal oxides},''
  \href{http://dx.doi.org/10.1103/PhysRevLett.98.146403}{{\em Physical Review
  Letters} {\bfseries {98}} no.~{14}, ({Apr 6}, {2007}) {146403}}.

\bibitem{Sawa:Review}
A.~Sawa, ``{Resistive switching in transition metal oxides},''
  \href{http://dx.doi.org/10.1016/S1369-7021(08)70119-6}{{\em {Materials
  Today}} {\bfseries {11}} no.~{6}, ({Jun}, {2008}) {28--36}}.

\bibitem{Shono:PCMO}
K.~Shono, H.~Kawano, T.~Yokota, and M.~Gomi, ``{Origin of negative differential
  resistance observed on bipolar resistance switching device with
  Ti/Pr$_{0.7}$Ca$_{0.3}$MnO$_3$/Pt structure},''
  \href{http://dx.doi.org/10.1143/APEX.1.055002}{{\em {Applied Physics
  Express}} {\bfseries {1}} no.~{5}, (May, {2008}) {055002}}.

\bibitem{Harada:PCMO}
T.~Harada, I.~Ohkubo, K.~Tsubouchi, H.~Kumigashira, T.~Ohnishi, M.~Lippmaa,
  Y.~Matsumoto, H.~Koinuma, and M.~Oshima, ``{Trap-controlled
  space-charge-limited current mechanism in resistance switching at
  Al/Pr$_{0.7}$Ca$_{0.3}$MnO$_3$ interface},''
  \href{http://dx.doi.org/10.1063/1.2938049}{{\em Applied Physics Letters}
  {\bfseries {92}} no.~{22}, ({Jun}, {2008}) {222113}}.

\bibitem{Odagawa}
A.~Odagawa, H.~Sato, I.~Inoue, H.~Akoh, M.~Kawasaki, Y.~Tokura, T.~Kanno, and
  H.~Adachi, ``{Colossal electroresistance of a Pr$_{0.7}$Ca$_{0.3}$MnO$_3$
  thin film at room temperature},''
  \href{http://dx.doi.org/10.1103/PhysRevB.70.224403}{{\em Physical Review B}
  {\bfseries {70}} no.~{22}, ({Dec}, {2004}) {224403}}.

\bibitem{Tokunaga:APL}
Y.~Tokunaga, Y.~Kaneko, J.~He, T.~Arima, A.~Sawa, T.~Fujii, M.~Kawasaki, and
  Y.~Tokura, ``{Colossal electroresistance effect at metal
  electrode/La$_{1-x}$Sr$_{1+x}$MnO$_4$ interfaces},''
  \href{http://dx.doi.org/10.1063/1.2208922}{{\em Applied Physics Letters}
  {\bfseries {88}} no.~{22}, ({May 29}, {2006}) {223507}}.

\bibitem{Asanuma:PCMO:switching}
S.~Asanuma, H.~Akoh, H.~Yamada, and A.~Sawa, ``{Relationship between resistive
  switching characteristics and band diagrams of Ti/Pr$_{1-x}$Ca$_x$MnO$_3$
  junctions},'' \href{http://dx.doi.org/10.1103/PhysRevB.80.235113}{{\em
  Physical Review B} {\bfseries {80}} no.~{23}, ({Dec}, {2009}) {235113}}.

\bibitem{Chua:Memristor}
L.~Chua, ``Memristor - the missing circuit element,''
  \href{http://dx.doi.org/10.1109/TCT.1971.1083337}{{\em {IEEE Transactions on
  Circuit Theory}} {\bfseries {CT18}} no.~{5}, ({1971}) {507--519}}.

\bibitem{Strukov:memristor}
D.~B. Strukov, G.~S. Snider, D.~R. Stewart, and R.~S. Williams, ``The missing
  memristor found,'' \href{http://dx.doi.org/10.1038/nature06932}{{\em Nature}
  {\bfseries 453} no.~7191, (2008) 80--83}.

\bibitem{Renner:BCMO}
C.~Renner, G.~Aeppli, B.~Kim, Y.~Soh, and S.~Cheong, ``{Atomic-scale images of
  charge ordering in a mixed-valence manganite},''
  \href{http://dx.doi.org/10.1038/416518a}{{\em {Nature}} {\bfseries {416}}
  no.~{6880}, ({Apr 4}, {2002}) {518--521}}.

\bibitem{Ma:LPCMO}
J.~Ma, D.~Gillaspie, E.~Plummer, and J.~Shen, ``{Visualization of localized
  holes in manganite thin films with atomic resolution},''
  \href{http://dx.doi.org/10.1103/PhysRevLett.95.237210}{{\em {Physical Review
  Letters}} {\bfseries {95}} no.~{23}, ({Dec 2}, {2005}) {237210}}.

\bibitem{Fuchigami}
K.~Fuchigami, Z.~Gai, T.~Z. Ward, L.~F. Yin, P.~C. Snijders, E.~W. Plummer, and
  J.~Shen, ``{Tunable metallicity of the La$_{5/8}$Ca$_{3/8}$MnO$_3$(001)
  surface by an oxygen overlayer},''
  \href{http://dx.doi.org/10.1103/PhysRevLett.102.066104}{{\em Physical Review
  Letters} {\bfseries {102}} no.~{6}, (Feb, {2009}) 066104}.

\bibitem{Roessler}
S.~R\"o{\ss}ler, B.~Padmanabhan, S.~Elizabeth, H.~L. Bhat, F.~Steglich, and
  S.~Wirth, ``{Atomically resolved scanning tunneling microscopy on perovskite
  manganite single crystals},'' \href{http://dx.doi.org/10.1063/1.3432753}{{\em
  Applied Physics Letters} {\bfseries {96}} no.~{20}, (May 17, {2010}) 202512}.

\bibitem{renner:LSMO}
H.~Ronnow, C.~Renner, G.~Aeppli, T.~Kimura, and Y.~Tokura, ``{Polarons and
  confinement of electronic motion to two dimensions in a layered manganite},''
  \href{http://dx.doi.org/10.1038/nature04650}{{\em {Nature}} {\bfseries {440}}
  no.~{7087}, ({Apr 20}, {2006}) {1025--1028}}.

\bibitem{Tokura:PSMO}
Y.~Tokunaga, T.~J. Sato, M.~Uchida, R.~Kumai, Y.~Matsui, T.~Arima, and
  Y.~Tokura, ``{Versatile and competing spin-charge-orbital orders in the
  bilayered manganite system Pr(Sr$_{1-y}$Ca$_y$)$_2$Mn$_2$O$_7$},''
  \href{http://dx.doi.org/10.1103/PhysRevB.77.064428}{{\em Physical Review B}
  {\bfseries {77}} no.~{6}, ({Feb}, {2008}) {064428}}.

\bibitem{Plummer2}
E.~Plummer, Ismail, R.~Matzdorf, A.~Melechko, and J.~Zhang, ``{The next 25
  years of surface physics},''
  \href{http://dx.doi.org/10.1016/S0079-6816(01)00014-4}{{\em Progress in
  Surface Science} {\bfseries {67}} no.~{1-8}, (June, {2001}) {17--44}}.

\bibitem{Takagi}
Y.~Takagi, K.~Hanai, H.~Hosokawa, H.~Ishibashi, T.~Ishikawa, A.~Saito,
  Y.~Kuwahara, and Y.~Taguchi, ``{Roughening Surface of Layered Manganite
  La$_{0.5}$Sr$_{1.5}$MnO$_4$ by Scanning Tunneling Microscopy},''
  \href{http://dx.doi.org/10.1143/JJAP.47.6456}{{\em Japanese Journal of
  Applied Physics} {\bfseries {47}} no.~{8, Part 1}, (Aug, {2008})
  {6456--6458}}.

\bibitem{Wakabayashi}
Y.~Wakabayashi, M.~H. Upton, S.~Grenier, J.~P. Hill, C.~S. Nelson, J.~W. Kim,
  P.~J. Ryan, A.~I. Goldman, H.~Zheng, and J.~F. Mitchell, ``{Surface effects
  on the orbital order in the single-layered manganite
  La$_{0.5}$Sr$_{1.5}$MnO$_4$},''
  \href{http://dx.doi.org/10.1038/nmat2061}{{\em Nature Materials} {\bfseries
  {6}} no.~{12}, (Dec, {2007}) {972--976}}.

\bibitem{Loviat}
F.~Loviat, H.~M. Ronnow, C.~Renner, G.~Aeppli, T.~Kimura, and Y.~Tokura, ``{The
  surface layer of cleaved bilayer manganites},''
  \href{http://dx.doi.org/10.1088/0957-4484/18/4/044020}{{\em {Nanotechnology}}
  {\bfseries {18}} no.~{4}, ({Jan 31}, {2007}) {044020}}.

\bibitem{Eigler:switch}
D.~M. Eigler, C.~Lutz, and W.~Rudge, ``An atomic switch realized with the
  scanning tunnelling microscope,''
  \href{http://dx.doi.org/10.1038/352600a0}{{\em Nature} {\bfseries {352}}
  no.~{6336}, ({Aug 15}, {1991}) {600--603}}.

\bibitem{Quaade:atomswitch}
U.~Quaade, K.~Stokbro, R.~Lin, and F.~Grey, ``{Single-atom reversible recording
  at room temperature},''
  \href{http://dx.doi.org/10.1088/0957-4484/12/3/311}{{\em Nanotechnology}
  {\bfseries {12}} no.~{3}, ({Sep}, {2001}) {265--272}}.

\end{thebibliography}
\end{document}